%% file: main.tex
\PassOptionsToPackage{table}{xcolor}
\documentclass[sigconf, nonacm]{acmart}
\usepackage{hyperref}
\usepackage{lineno}
\usepackage{hyperxmp}
\usepackage{alltt}
\usepackage{listings}

\usepackage{tcolorbox} 
\usepackage[linesnumbered,ruled,vlined]{algorithm2e}
\usepackage{amsmath}
\usepackage{listings}
\usepackage{geometry}
\usepackage[linesnumbered,ruled,vlined]{algorithm2e}
\usepackage{multicol}
\usepackage{lipsum}
\usepackage{mdframed}
\usepackage{tikz}
\usepackage{pgf-pie}
\usepackage{tcolorbox}
\usepackage{amsmath}
\usepackage{mdframed}

\newmdenv[
    linecolor=gray!40,
    backgroundcolor=gray!5,
    roundcorner=2pt,
    leftmargin=0pt,
    rightmargin=0pt,
    innerleftmargin=10pt,
    innerrightmargin=10pt,
    innertopmargin=10pt,
    innerbottommargin=10pt,
    linewidth=1pt
]{definitionframe}

\definecolor{codegreen}{rgb}{0,0.6,0}
\definecolor{codegray}{rgb}{0.5,0.5,0.5}
\definecolor{codeblue}{rgb}{0,0,1}
\definecolor{backcolour}{rgb}{0.95,0.95,0.96}

\lstdefinestyle{mystyle}{
    backgroundcolor=\color{backcolour},   
    commentstyle=\color{codegreen},
    keywordstyle=\color{codeblue},
    numberstyle=\tiny\color{codegray},
    stringstyle=\color{codegreen},
    basicstyle=\ttfamily\footnotesize\linespread{0.9},  
    breakatwhitespace=false,         
    breaklines=true,                 
    captionpos=b,                    
    keepspaces=true,                 
    numbers=left,
    framerule=1.4pt,
    numbersep=5pt,    
    frame=single, 
    showspaces=false,                
    showstringspaces=false,
    showtabs=false,                  
    tabsize=2,
    framexleftmargin=10pt, 
    xleftmargin=13pt, 
}

\usepackage{caption}

\AtBeginDocument{%
  \providecommand\BibTeX{{%
    \normalfont B\kern-0.5em{\scshape i\kern-0.25em b}\kern-0.8em\TeX}}}

\usepackage{enumitem}
\usepackage{float}
\usepackage{soul}
\usepackage{subcaption}

\sethlcolor{white}

\newif\ifshowblue
\showbluefalse   

\newcommand{\blueadd}[1]{%
  \ifshowblue
    \textcolor{blue}{#1}%
  \else
    #1%
  \fi
}

\setcopyright{none}
\acmPrice{}
\makeatletter
\let\ACM@cc@type\@empty
\makeatother
\begin{document}
\title[Improving LLM-based Code Evaluation with Question-Specific Rubrics]{Rubric Is All You Need: Improving LLM-based Code Evaluation With Question-Specific Rubrics}

    \author{Aditya Pathak}
    \authornote{These authors contributed equally to this work.}
    \affiliation{
        \institution{BITS Pilani}
        \city{Pilani}
        \country{India}
    }

    \author{Rachit Gandhi}
    \authornotemark[1]
    \affiliation{
        \institution{BITS Pilani}
        \city{Pilani}
        \country{India}
    }
    
    \author{Vaibhav Uttam}
    \authornotemark[1]
    \affiliation{
        \institution{BITS Pilani}
        \city{Pilani}
        \country{India}
    }

    \author{Arnav Ramamoorthy}
    \authornotemark[1]
    \affiliation{
        \institution{BITS Pilani}
        \city{Pilani}
        \country{India}
    }

    \author{Pratyush Ghosh}
    \authornotemark[1]
    \affiliation{
        \institution{BITS Pilani}
        \city{Pilani}
        \country{India}
    }

    \author{Aaryan Raj Jindal}
    \affiliation{
        \institution{BITS Pilani}
        \city{Pilani}
        \country{India}
    }

    \author{Shreyash Verma}
    \affiliation{
        \institution{BITS Pilani}
        \city{Pilani}
        \country{India}
    }
    
    \author{Aditya Mittal}
    \affiliation{
        \institution{BITS Pilani}
        \city{Pilani}
        \country{India}
    }
    
    \author{Aashna Ased}
    \affiliation{
        \institution{BITS Pilani}
        \city{Pilani}
        \country{India}
    }
    
    \author{Chirag Khatri}
    \affiliation{
        \institution{BITS Pilani}
        \city{Pilani}
        \country{India}
    }

    \author{Yashwanth Nakka}
    \affiliation{
        \institution{BITS Pilani}
        \city{Pilani}
        \country{India}
    }
    
    \author{Devansh}
    \affiliation{
        \institution{BITS Pilani}
        \city{Pilani}
        \country{India}
    }
    
    \author{Jagat Sesh Challa}
    \affiliation{
        \institution{BITS Pilani}
        \city{Pilani}
        \country{India}
    }
    
    \author{Dhruv Kumar}
    \affiliation{
        \institution{BITS Pilani}
        \city{Pilani}
        \country{India}
    }


\renewcommand{\shortauthors}{Aditya Pathak et al.}

\begin{abstract}
\input{files/00-abstract}
\end{abstract}

\ccsdesc[500]{Social and professional topics~Student assessment}
\ccsdesc[500]{Computing methodologies~Natural language processing}

\keywords{Large Language Models, Code Assessment and Grading}



\maketitle

\section{Introduction}\label{sec:intro}
\input{files/01-introduction}

\section{Related Work}\label{sec:related_work}
\input{files/02-related_work}

\section{Dataset}\label{sec:dataset}
\input{files/03a-dataset}

\section{Proposed Techniques}\label{sec:method}
\input{files/03-methodology}

\section{Metrics and Evaluation}\label{sec:evaluation}
\input{files/04-Evaluation}
\vspace{-1em}

\section{Results}\label{sec:results}
\input{files/04a-results}
\input{files/05-discussion}
\section{Discussion}\label{sec:discussion}
\input{files/06b-discussion}
\section{Limitations and Future Work}\label{sec:limitations}
\input{files/06a-limitations}
\section{Conclusion}\label{sec:conclusion}

\input{files/06-conclusion}
\section{Acknowledgments}
\input{files/061-acknowledgements}

\appendix
\input{files/07-appendix}
\bibliographystyle{ACM-Reference-Format}

\bibliography{chatgpt-1}


\end{document}

%% file: files/00-abstract.tex

Since the emergence of Large Language Models (LLMs) popularized by the release of GPT-3 and ChatGPT, LLMs have shown remarkable promise in programming-related tasks. While code generation using LLMs has become a popular field of research, code evaluation using LLMs remains under-explored. In this paper, we focus on LLM-based code evaluation and attempt to fill in the existing gaps. We propose multi-agentic novel approaches using \emph{question-specific rubrics} tailored to the problem statement, arguing that these perform better for logical assessment than the existing approaches that use \emph{question-agnostic rubrics}. To address the lack of suitable evaluation datasets, we introduce two datasets: a Data Structures and Algorithms dataset containing 150 student submissions from a popular Data Structures and Algorithms practice website, and an Object Oriented Programming dataset comprising 80 student submissions from undergraduate computer science courses. In addition to using standard metrics (Spearman Correlation, Cohen's Kappa), we additionally propose a new metric called as Leniency, which quantifies evaluation strictness relative to expert assessment. Our comprehensive analysis demonstrates that \emph{question-specific rubrics} significantly enhance logical assessment of code in educational settings, providing better feedback aligned with instructional goals beyond mere syntactic correctness.

%% file: files/01-introduction.tex

The integration of Large Language Models (LLMs) into computing education has led to groundbreaking advancements, transforming both students and educators experiences \cite{hellas2024experiencesintegratinglargelanguage, joshi2023withgreatpowercomes, hicke2023aitaintelligentquestionanswerteaching, raihan2024largelanguagemodelscomputer} . In computing education particularly, LLMs have demonstrated potential in generating instructional content \cite{Sarsa_2022, phung2023generativeaiprogrammingeducation}, providing personalized tutoring and doubt solving \cite{liffiton2023codehelpusinglargelanguage, kotalwar2025hintsinbrowserbenchmarkinglanguagemodels} as well as assisting with code evaluation and grading \cite{10.1145/3626252.3630803, Alkafaween_2024, phung2023generativeaiprogrammingeducation}.
Despite these advancements, there are significant gaps in using LLMs for code evaluation and grading. Most studies focus on generating feedback, helping students identify errors and improve their work \cite{Alkafaween_2024, pankiewicz2023largelanguagemodelsgpt, 10.1145/3491101.3519665}. Existing studies have demonstrated LLMs' ability to provide meaningful insights into code quality, syntax, and logic \cite{pankiewicz2023largelanguagemodelsgpt, anishka2024chatgptplayroleteaching}. Yet, grading, a critical and labor-intensive responsibility for instructors \cite{article_grading_time, NORMANN2023104336}, has received comparatively little attention \cite{kiesler2023exploringpotentiallargelanguage,Jacobs_2024}. The evaluation and grading of student code is a critical component of computing education, as it helps instructors assess students' understanding of programming concepts and computational logic. This process consumes significant time and energy for instructors \cite{article_grading_time, NORMANN2023104336}, making it a prime candidate for automation.

Automated grading systems have been in use for a while \cite{10.1145/1930464.1930480, Messer_2024}, but their effectiveness has often been limited due to their reliance on rigid test cases and simplistic evaluation criteria. Prior work in the domain of automated grading has focused on either introductory programming courses or on short answer evaluation \cite{raihan2024largelanguagemodelscomputer, Alkafaween_2024, Yousef2024}. Moreover, we note that existing work on feedback and grading of programming assignments has focused on \emph{question-agnostic (QA) rubrics} \cite{phung2023generativeaiprogrammingeducation, fan2025sedarevalautomatedevaluationusing}, emphasizing generic criteria such as correctness and syntax across diverse problems, but in reality we find that the human instructors actually use \emph{question-specific (QS) rubrics}.  

In this paper, we focus on evaluating the effectiveness of LLMs in grading student code using \emph{question-specific rubrics} in more advanced computing courses, such as Object-Oriented Programming (OOP) and Data Structures and Algorithms (DSA), where complex problem-solving demands greater precision and context-awareness. We hypothesize that such an approach utilizing \textbf{question-specific rubrics} would yield more accurate evaluations compared to question-agnostic methods. Thus, the primary research question we seek to answer in this paper is:

\textbf{\textit{"How effectively can an LLM-based grader evaluate student code using a question-specific rubric compared to a question-agnostic rubric? 
What measurable differences exist in evaluation quality and feedback specificity between these two approaches?"}}



We propose three novel techniques for code evaluation: (1) \textbf{Complete Rubric Evaluation (CRE)}, is a LLM-agent which assesses student submissions against the entire rubric, prioritizing logical correctness and intentionally overlooking syntax errors, to focus primarily on conceptual understanding. A deterministic compiler-equipped agent is used for checking syntactical correctness ; (2) \textbf{Pointwise Rubric Evaluation (PRE)}, is similar to CRE, but evaluates submissions by individually checking each criterion within the rubric, providing detailed and granular feedback. It is comparatively more resource-intensive; and (3) \textbf{Ensembling Method Evaluation (EME)}, which enhances reliability by aggregating rubric-based evaluations through majority voting and related ensemble mechanisms, ensuring robust feedback and assessment consistency. Additionally, we introduce a new evaluation metric called \textbf{Leniency}, which measures the strictness or leniency of automated evaluations compared to expert human assessments, providing insights into the relative evaluation rigor.

Our empirical findings demonstrate that question-specific rubrics substantially outperform question-agnostic rubrics, leading to improved accuracy, feedback relevance, and alignment with educational objectives.

Conducting this research required a suitable dataset containing student code submissions, model solutions, grading rubrics  and feedback. We found that no such dataset was available publicly. To fill this gap, we created our own dataset, drawing from student submissions for OOP and DSA programming exercises. This dataset includes: (1) problem description, (2) student-submitted code, (3) model solutions, (4) grading rubrics, and (5) qualitative feedback, all of which are essential for benchmarking the performance of LLMs in code evaluation tasks. We will be releasing this dataset soon to the public to facilitate further research in this area. 
The main research contributions of this work are as follows:
\begin{itemize}
\item We present a new dataset that comprises submissions from two important courses in undergraduate computing education: (1) Object-Oriented Programming (OOP), and (2) Data Structures and Algorithms (DSA). (\textbf{\S \ref{sec:dataset}})
\item We introduce three novel techniques for code evaluation and grading: (1) Complete Rubric Evaluation (CRE), (2) Pointwise Rubric Evaluation (PRE) 
and (3) Ensembling Method Evaluation (EME). (\textbf{\S \ref{sec:method}})
\item We present a new metric that measures the strictness or leniency of an evaluation system based on rubric. 
(\textbf{\S \ref{sec:evaluation}})
\item We perform a comprehensive evaluation of the proposed techniques showing that our proposed techniques outperform all other techniques on both the datasets, achieving high correlation with human graders. (\textbf{\S \ref{sec:results}})

\end{itemize}


By demonstrating the effectiveness of question-specific rubrics in LLM-based grading systems, we pave the way for more accurate and efficient automated evaluation tools. These tools have the potential to save instructors significant time and effort, allowing them to focus on other aspects of teaching and mentoring. Additionally, students stand to benefit from more detailed and contextually relevant feedback, which can help them identify areas for improvement and deepen their understanding of programming concepts. We have made our dataset and code publicly available on HuggingFace\footnote{https://huggingface.co/datasets/BITS-Pilani-GRC/RubricEval} and GitHub\footnote{https://github.com/BITS-Pilani-GRC/Rubric-Grader} respectively.

%% file: files/02-related_work.tex
Effective assessment and feedback mechanisms are fundamental to programming education, serving as critical scaffolds that allow for guiding students through the problem-solving process with proper guidance. Prather et al. \cite{10.1145/3105726.3106169} emphasize that effective feedback extending beyond the binary correct/incorrect judgments based on test-cases,  generally provided by traditional automated grading systems  is crucial for fostering self-regulated learning. Similarly, Hao and Tsikerdekis \cite{9028686} note that well-designed feedback promotes metacognitive development. The increasing enrollment in programming courses has rendered grading the programming assignments increasingly labor-intensive and time consuming \cite{8802114}. To address this challenge, instructors frequently rely on automated grading tools \cite{10.1145/3231711}.

\begin{table*}[t]
\centering
\begin{tabular}{lcccccc}
\hline
\textbf{Technique} & \textbf{Grading} & \textbf{Feedback} & \textbf{\begin{tabular}[c]{@{}c@{}}Programming\\ Focused\end{tabular}} & \textbf{\begin{tabular}[c]{@{}c@{}}Question\\ Agnostic\\ Rubric\end{tabular}} & \textbf{\begin{tabular}[c]{@{}c@{}}Question\\ Specific\\ Rubric\end{tabular}} \\
\hline
BLEU \cite{papineni-etal-2002-bleu} & \checkmark & $\times$ & $\times$ & $\times$ & $\times$ \\
CodeBLEU \cite{ren2020codebleumethodautomaticevaluation} & \checkmark & $\times$ & \checkmark & $\times$ & $\times$ \\
CodeBERTScore \cite{zhou2023codebertscoreevaluatingcodegeneration} & \checkmark & $\times$ & \checkmark & $\times$ & $\times$ \\
ICE-Score \cite{zhuo2024icescoreinstructinglargelanguage} & \checkmark & $\times$ & \checkmark & $\times$ & $\times$ \\
CodeJudge \cite{tong2024codejudgeevaluatingcodegeneration} & \checkmark & \checkmark & \checkmark & $\times$ & $\times$ \\
\begin{tabular}[c]{@{}l@{}}\\\end{tabular} Phung et al\cite{phung2023generativeaiprogrammingeducation} & \checkmark & \checkmark & \checkmark & \checkmark & $\times$ \\
CodEv \cite{Tseng_2024} & \checkmark & \checkmark & \checkmark & \checkmark & $\times$ & \\
\textbf{Proposed Techniques}  & \checkmark & \checkmark & \checkmark & \checkmark & \checkmark \\
\hline
\end{tabular}
\caption{Comparison of LLM-Based Techniques for Grading and Feedback}
\label{tab:code-assessment}
\end{table*}


\subsection{Limitations of Traditional Auto-grading Approaches}
Conventional autograding systems have primarily relied on predefined test cases to evaluate student submissions. Lobb et al. \cite{10.1145/2810041} developed a system that executed student code and compared outputs against expected results derived from instructor-defined test cases. While this approach allowed for basic assessment of functional correctness, 
these systems often lack the ability to understand underlying semantic errors or provide nuanced explanations \cite{10.1145/2899415.2899422}.

Additional limitation of test-case-based approaches is the burden placed on instructors to create comprehensive test suites manually. Keuning et al \cite{10.1145/2899415.2899422} highlighted the challenges associated with generating test cases that effectively cover edge cases and potential error conditions. Furthermore, 
traditional systems often focus exclusively on functional correctness, neglecting aspects such as code style, efficiency, and adherence to best practices \cite{10.5555/2541917.2541921}. Finally, test suites require successful compilation where even minor syntactical errors can render logically correct code non-functional. This results in an incomplete assessment of student programming abilities and fails to promote holistic coding skills development.

Our approach addresses these challenges by employing instructor-defined rubrics that provide specific evaluation criteria. This enables more meaningful feedback beyond binary judgments while maintaining scalability. Though rubrics are also manually intensive, they have great upside in helping instructors evaluate students consistently and objectively \cite{Chowdhury2018ApplicationOR}, while providing students with clear expectations and constructive feedback to identify their strengths and weaknesses \cite{8820850}.

\subsection{Emergence of LLMs in Programming Assessment}

The emergence of Large Language Models (LLMs) marks a significant paradigm shift in autograding approaches. Recent studies by Denny et al. \cite{10.1145/3641554.3701791} have outlined how LLMs can leverage their extensive training data to grasp nuances of code, identify common errors, and explain complex concepts in an accessible manner.
Multiple studies \cite{phung2023generativeaiprogrammingeducation, Azaiz_2024, kiesler2023exploringpotentiallargelanguage} have shown that advanced LLMs such as GPT-4 \cite{openai_gpt-4_2023} can approach human-level feedback quality and deliver formative insights beyond simple correctness checks. Furthermore, Leinonen et al. \cite{Hellas_2023} analyzed techniques for using LLMs to improve programming error messages, making them more informative and actionable for students.

Though the above work highlights the potential of LLMs in the field, many studies also raised questions on the reliability and consistency of LLM-based evaluation \cite{app15020671}. While the majority of the existing work focuses purely on providing effective feedback, we comprehensively also cover grading the assignments which is essential for both instructors and students, especially for a large-scale programming course. Our approach using question-specific rubrics also ensures structured evaluation criteria that combine human expertise with AI capabilities to provide consistent, contextual feedback.

\subsection{LLM-based Techniques for Grading and Feedback}

\autoref{tab:code-assessment} compares the major LLM-based approaches for programming assessment. 
CodeBERTScore \cite{zhou2023codebertscoreevaluatingcodegeneration} leverages pre-trained BERT \cite{devlin2019bertpretrainingdeepbidirectional} models to encode semantic vectors of reference and generated code, measuring the similarity between these vectors rather than relying on token-matching approaches. This method improved upon earlier techniques like traditional BLEU \cite{papineni-etal-2002-bleu} (which treated code as mere text) and its specialized derivative CodeBLEU \cite{ren2020codebleumethodautomaticevaluation} (which incorporated weighted n-gram matching, AST comparison, and data-flow analysis). Despite CodeBERTScore's innovations in capturing semantic meaning, it still faces a fundamental limitation: context similarity doesn't necessarily represent semantic similarity, resulting in suboptimal performance when evaluating functionally identical code implemented with different approaches which is a particular challenge in educational environments.

ICE-Score \cite{zhuo2024icescoreinstructinglargelanguage} has explored using LLMs directly for code evaluation without relying on test cases. However, this approach demonstrates limited correlation with human judgment and remains susceptible to LLM hallucinations, particularly when evaluating complex code with intricate semantics. Similarly, CodeJudge \cite{tong2024codejudgeevaluatingcodegeneration} leverages "slow thinking" to guide LLMs in evaluating code semantics. By decomposing evaluation into step-by-step analysis followed by summarization, and introducing a taxonomy of code inconsistencies with severity levels, it achieves better correlation with semantic correctness compared to other methods. 
The above mentioned techniques are primarily for benchmarking the correctness and quality of code generated by LLMs but are unable to generate any meaningful feedback for student code submissions.



Beyond these approaches, we examined rubric-based techniques for code evaluation and identified a significant gap in research exploring such methods. 
Although rubric creation is as time-intensive for instructors as building test suites, the resulting rubrics provide far greater educational value to students. Unlike test suites, which provide only error-based feedback and do not work when the student code does not compile, rubrics deliver direct, structured feedback that systematically supports both learning and assessment. 
For ease of our discussion, we further subdivide rubric-based programming evaluation on the basis of rubric specificity: \textbf{(1) question-agnostic rubric} and \textbf{(2) question-specific rubric}. These are formally defined in \autoref{fig:rubric-definitions}. 

\begin{figure*}[t]
\centering
\begin{definitionframe}
\textbf{Rubric Types for Programming Problems: }
For a programming problem $P$ with description $D$ and expected solutions $S$), we define:
\begin{itemize}
    \item A \textbf{question-agnostic rubric} $R_{\text{agnostic}}$ as a set of evaluation criteria independent of $ specific-logical $ $requirements $. 
    Formally, $R_{\text{agnostic}}(P_1) = R_{\text{agnostic}}(P_2)$ for any distinct problems $P_1, P_2$. These criteria often cover general aspects like code style, basic correctness, and efficiency.
    \item A \textbf{question-specific rubric} $R_{\text{specific}}(P)$ as a set of evaluation criteria tailored to $ unique $ $ logical-requirements $  and $ constraints$. Formally, $R_{\text{specific}}(P_1) \neq R_{\text{specific}}(P_2)$ for distinct problems $P_1, P_2$, with criteria directly referencing $D$ and $S$.
\end{itemize}
\end{definitionframe}
\caption{Formal definitions of rubric types used in LLM-based code evaluation.}
\label{fig:rubric-definitions}
\end{figure*}

LLM-based techniques using question-agnostic rubrics often fail to capture the nuances of specific programming problems, leading to misaligned evaluations, and still struggle with grading feedback compared to human tutors \cite{phung2023generativeaiprogrammingeducation}. Phung et al. \cite{phung2023generativeaiprogrammingeducation} showed that both GPT-4 and ChatGPT frequently misidentify code issues and incorrectly assign points for general correctness and edge cases when using general rubrics, suggesting question-specific rubrics might be more effective for automated assessment systems. More recent developments include domain-specific approaches like CodEV \cite{Tseng_2024}, which leverage LLMs, Chain of Thought, LLM ensembles, and a question-agnostic rubric to improve score accuracy and consistency.

Xie et al. \cite{xie2024gradelikehumanrethinking} propose a multi-agent “Grade Like a Human” system which creates context-aware rubrics for short-answer questions. It then uses these rubrics to score responses, give tailored feedback, and run accuracy checks. Because it targets short-answer grading, this approach does not transfer directly to programming assessments.

To the best of our knowledge, this study is the first to utilize question-specific rubrics for grading and providing feedback on student code submissions. Our method distinctly separates the assessment of logical reasoning from syntactic correctness, mirroring real-world practice where conceptual understanding is given preference over code syntax.


%% file: files/03a-dataset.tex
We present two datasets, one based on Object Oriented Programming (OOP) and second based on Data Structures and Algorithms (DSA). These courses are taken by all Computer Science undergraduates making them relevant to our study. Overall, our dataset contains 230 student submissions (80 from OOP and 150 from DSA). Full Dataset is available on HuggingFace \footnote{https://huggingface.co/datasets/BITS-Pilani-GRC/RubricEval}. Below, we provide the high level details about the dataset:

\subsection{ OOP Dataset Construction}

For the OOP dataset, we rely on student submissions as part of a programming exam conducted in an OOP course at BITS Pilani in Fall 2024. 
The programming exam consisted of one Java programming question which further consisted of seven methods which needed to be implemented by the students. Students were provided with a structured starter template that included class declarations, helper functions, and predefined method scopes for implementation. A main function was also supplied to the students for testing and validating their solutions. The tasks in these methods focussed on applying object-oriented programming principles to handle file I/O, data filtering, and data updates.

\noindent
\begin{minipage}{0.50\textwidth}
\begin{tcolorbox}[
    colback=gray!10,    
    colframe=black,     
    sharp corners,      
    width=\textwidth,   
    before skip=0pt,
    left=3pt
    ]
\begin{alltt}\footnotesize
CricketDataHandler: readPlayersFromFile Method [9 marks] 
Write code for reading player data from the input CSV 
file and creating 
a list of Player objects.
• Step 1: Create an empty list to store player details.[1 mark]
• Step 2: Open the specified file for reading data. [1 mark]
• Step 3: Ignore the first line since it contains the 
  column names. [1 mark]
• Step 4: Read each line one by one until reaching the end of the 
  file. [1 mark]
• Step 5: Split the line into different pieces of information.
  [1 mark]
• Step 6: Create a new player using this information.[1 mark]
• Step 7: Add the new player to the list. [1 mark]
• Step 8: Close the file after reading all data. [1 mark]
• Step 9: Return the complete list of players. [1 mark]
\end{alltt}
\end{tcolorbox}
\end{minipage} 
\hfill
\begin{minipage}{0.49\textwidth}
\begin{lstlisting}[language=Java,label={lst:stud_code}, mathescape=true]
public List<player> readPlayersFromFile(String fileName) throwsIOException {
List<player> players=new List<player>();
Scanner sc= null;
Sc = new Scanner(new FileInputStream("InputCricketData.csv"));
sc.nextLine();
while(sc.hasNext()){
    Player p = new Player();
    String a=sc.nextLine();
    String s1=a.split(",")[0];
    String s2=a.split(",")[1];
    String s3=a.split(",")[2];
    String s4=a.split(",")[3];
    String s5=a.split(",")[4];
    p.setPlayerName(s1);
    p.setRole(s2);
    p.setRunsScored(Integer.parseInt(s3));
    p.setWickets Taken (Integer.parseInt(s4));
    p.setTeamName(s5);
    players.add(p);}
    sc.close();
    return players;}
\end{lstlisting}
\end{minipage}
\vspace{-1em}  
\captionof{figure}{Sample Question from OOP dataset and corresponding student submission}
\label{fig:oop-sample}
\vspace{1em} 

We selected 80 student submissions from a total of 350 submissions, all from undergraduate sophomores. The submissions were graded by Teaching Assistants (TAs) and categorized into four score ranges: 0–10, 10–20, 20–30, and 30–35 marks (max score was 35). From each category, 20 solutions were selected at random, totaling to 80 student submissions. The dataset consists of the following components:
\begin{enumerate}[leftmargin=*]
    \item \textbf{Detailed Problem Statement:}
 A detailed problem statement provided by the instructor, outlining clear, step-by-step instructions for implementing each function in every part of the question.
 \item \textbf{Scaffold Code:}
The code provides a foundational class structure with templates and clear markers indicating where students should insert their code.

\item \textbf{Rubric:}
 The rubric outlines each step as a distinct component, assigned either 1 or 2 marks, with no partial grading. A 2-mark step is awarded either 0 or 2 marks. Each of the seven subproblems follows a structured sequence, with steps arranged in a logical order based on their implementation in the codebase.

\item \textbf{Model Solution: }
A model solution prepared by the instructor, which includes implementations of all steps while strictly adhering to the guidelines in the problem statement and rubric.

\item \textbf{Student Submissions:}
Each student submission consists of three parts:
\begin{itemize}[leftmargin=*]
    \item \textbf{Student code:} The student’s submitted Java code.
    \item \textbf{Grades:} Part-by-part evaluation conducted through a consensus-driven approach by two graders.
    \item \textbf{Feedback:} For each of the seven subproblems, a comprehensive feedback was created collaboratively by both graders based on the problem statement and rubric. Both graders are final-year students with extensive experience in programming courses, ensuring a thorough and consistent assessment process.   
\end{itemize}
\end{enumerate}

For example, in the student code in Figure \ref{fig:oop-sample}, in line 2, List<Player> players = new List<Player>(); is incorrect, as List is an interface and cannot be instantiated directly (Step 1). However, all subsequent steps are correct. The final score, determined by human annotators, is 8 out of 9, with individual step-wise marks assigned as 0, 1, 1, 1, 1, 1, 1, 1, 1.
\subsection{ DSA Dataset Construction}
To ensure diversity and comprehensiveness, we selected DSA problems from the Geeks for Geeks (GFG) practice website [7], spanning 9 topics and 3 difficulty levels (easy, medium, and hard). The distribution of
problems across topics and difficulty levels is shown in Figure \ref{tab:problem_distribution} and Figure \ref{fig:difficulty_distribution}.
For each problem, the dataset contains:
\begin{enumerate}
    \item \textbf{Problem Statement} - The problem statements were sourced from the descriptions provided on the website for each question. These include textual description of the problem, input size constraints and example input-output pairs to illustrate expected behavior.
    \item \textbf{Model Solution} - One model solution was taken from the editorial given with the problem.
    \item \textbf{Submissions} - Six distinct submissions were selected for each problem, representing different categories of outcomes : Correct (3 solutions), Wrong (1 solution), TLE (1 solution) and Compilation error (1 solution).
    \item \textbf{Rubric} - The rubric was designed by considering different approaches to solving the question. Marks were assigned to each step based on its relevance and significance across various possible solutions.
    \item \textbf{Feedback and Marking} - Two human graders collaboratively evaluated six different solutions per problem, following the predefined rubric. They also provided detailed feedback for each implementation step. Both graders, as final-year students with extensive programming experience, ensured a thorough and consistent assessment process.
\end{enumerate}

\begin{figure}[htbp]
\centering
\begin{minipage}{0.48\textwidth}
    \centering
    \begin{tabular}{lc}
    \toprule
    \textbf{Topic} & \textbf{Number of questions} \\
    \midrule
    Arrays & 4 \\
    Binary Search & 3  \\
    Bit Magic (Bitwise Operators) & 2  \\
    Dynamic Programming & 2  \\
    Graphs & 3  \\
    Hash & 3  \\
    Linked Lists & 2  \\
    Strings & 4  \\
    Trees & 2  \\
    \bottomrule
    \end{tabular}
    \caption{Distribution of Problems by Topic}\label{tab_dis}
    \label{tab:problem_distribution}
\end{minipage}%
\hfill
\begin{minipage}{0.48\textwidth}
    \centering
    \begin{tikzpicture}
    \pie[
        text=legend,
        radius=2,
        explode=0.1,
        color={green!60, orange!70, red!60}
    ]{
        48/Easy (12),
        32/Medium (8),
        20/Hard (5)
    }
    \end{tikzpicture}
    \caption{Distribution of Problems by Difficulty Level}
    \label{fig:difficulty_distribution}
\end{minipage}
\end{figure}

\blueadd{A sample comprising  student solution, problem statement, rubric, and grade \& feedback is shared in \autoref{fig:dsa-solution}, \autoref{fig:Sample_problem_statement}, \autoref{fig:Sample_rubric_statement},  and \autoref{fig:sample_feedback} respectively.}

\begin{lstlisting}[language=Java,label={lst:code}, mathescape=true]
class Solution {
    // Function is to check whether two strings are anagram of each other or not. 
    public static boolean areAnagrams (String s1, String s2) {
        // Your code here
        if(s1.length() != s2.length() ) 
            return false;
        int charcount[] = new int[256];
        for(int i = 0; i < s1.length(); i++ ){
            charCount[s1.charAt(i)]++;
            charCount[s2.charAt(i)]--;
        }
        for(int count: charCount)
            if( count != 0)
                return false;
        return true;
    }
}
\end{lstlisting}
\vspace{-\baselineskip}  
\captionof{figure}{Sample Student Solution for DSA}
\label{fig:dsa-solution}

\begin{figure*}[h]
\centering
\begin{tcolorbox}[
    colback=gray!20,    
    colframe=black,     
    sharp corners,      
    width=\textwidth,   
    ]

\begin{alltt}\small

Given two strings s1 and s2 consisting of lowercase characters. The task is to check whether two given 
strings are an anagram of each other or not. An anagram of a string is another string that contains the same 
characters, only the order of characters can be different. For example, "act" and "tac" are an anagram of
each other. Strings s1 and s2 can only contain lowercase alphabets.
Note: You can assume both the strings s1 & s2 are non-empty.
Examples:
Input: s1 = "geeks", s2 = "kseeg"
Output: true
Explanation: Both the string have same characters with same frequency. So, they are anagrams.
Input: s1 = "allergy", s2 = "allergic"
Output: false
Explanation: Characters in both the strings are not same, so they are not anagrams.
Input: s1 = "g", s2 = "g"
Output: true
Explanation: Character in both the strings are same, so they are anagrams.
Constraints: 1\ensuremath{\leq} s1.size(), s2.size() \ensuremath{\leq} 105
\end{alltt}
\end{tcolorbox}
\vspace{-\baselineskip}  
\captionof{figure}{Sample Problem Statement for DSA}
\label{fig:Sample_problem_statement}
\end{figure*}
\begin{figure*}[h]
\centering

\begin{tcolorbox}[
    colback=gray!20,    
    colframe=black,     
    sharp corners,      
    width=\textwidth,   
    ]
\begin{alltt}\small
Anagram:
1. Check if two Strings are Anagrams of each other
2. Solution 1:
\hspace{0.3cm}1. Initialize a map or dictionary or array to store the character frequencies. Initially, the frequency 
\hspace{0.5cm} for each character should be zero.[1 mark]
\hspace{0.3cm}2. For each character in the first string, the corresponding frequency is incremented by 1. [1 mark]
\hspace{0.3cm}3. For each character in the second string, decrement its corresponding frequency by 1. [1 mark]
\hspace{0.3cm}4. Iterate through the entire map or dictionary or array. If any frequency is non-zero, then return false. Else return 
    true. [1 mark]
3. Solution 2:
\hspace{0.3cm}1. Initialize a map or dictionary or array to store the character frequencies. Initially, the frequency 
\hspace{0.5cm} for each character should be zero. Keep separate data structures for each of the strings. [1 mark]
\hspace{0.3cm}2. For each character in the first string, the corresponding frequency in its data structure is 
\hspace{0.5cm} incremented by 1. [1 mark]
\hspace{0.3cm}3. For each character in the second string, the corresponding frequency in its data structure is 
\hspace{0.5cm} incremented by 1. [1 mark]
\hspace{0.3cm}4. Iterate through both the data structures. If the frequency for any unique character in both the 
\hspace{0.5cm} data structures does not match, return false. Else return true. [1 mark]
4. Solution 3:
\hspace{0.3cm}1. Sort both the strings. [2 marks]
\hspace{0.3cm}2. Return true if both the sorted strings are exactly same. Else, return false. [2 marks]

\end{alltt}
\end{tcolorbox}
\vspace{-\baselineskip}  
\captionof{figure}{Sample Rubric for DSA}
\label{fig:Sample_rubric_statement}
\newpage
\end{figure*}

\begin{figure*}[h]
\centering
\begin{tcolorbox}[
    colback=gray!20,    
    colframe=black,     
    sharp corners,      
    width=\textwidth,   
    ]
\begin{alltt}\small
"Selected rubric" Solution 1
"Feedback with marks"
1. Initialize a map or dictionary or array to store the character frequencies. Initially, the frequency 
\hspace{0.3cm} for each character should be zero. [1 mark]
\hspace{0.6cm}-The student initializes an array `charCount` to store character frequencies, which is correct.[1 mark]
2. For each character in the first string, the corresponding frequency is incremented by 1.[1 mark] 
\hspace{0.6cm}-The student correctly increments the frequency for each character in the first string `s1` using 
\hspace{0.6cm} `charCount[s1.charAt(i)]+`. [1 mark]
3. For each character in the second string, decrement its corresponding frequency by 1. [1 mark] 
\hspace{0.6cm}-The student correctly decrements the frequency for each character in the second string 's2' using 
\hspace{0.6cm} 'charCount[s2.charAt(i)]--'. [1 mark]
4. Iterate through the entire map or dictionary or array. If any frequency is non-zero, then return false. 
\hspace{0.3cm} Else return true. [1 mark]
\hspace{0.6cm}-The student iterates through the `charCount array and checks if any frequency is non-zero, returning 
\hspace{0.6cm} false if so, and true otherwise. [1 mark]
"Marks Scored out of total marks"
4 out of 4 marks

\end{alltt}
\end{tcolorbox}
\vspace{-\baselineskip}  
\captionof{figure}{Sample Grade and Feedback for DSA}
\label{fig:sample_feedback}
\end{figure*}

%% file: files/03-methodology.tex
\subsection{\textbf{Motivation}}
The proposed techniques discussed below were designed in order to evaluate submissions using question-specific rubrics. As discussed earlier, question-specific rubrics were introduced in order to emulate college or university-level grading of student code. Akin to college grading, we break down the problem statement into steps that the student is expected to implement. Marks are awarded with respect to the student's approach for each specific step. The marking is binary, so a student is awarded either full or zero marks for a step. Our datasets attempt to emulate a similar level of leniency and specificity with grading of the ground truth values.

\subsection{\textbf{Complete Rubric Evaluation (CRE)}}
\begin{figure*}
    \centering
    \includegraphics[width=\linewidth]{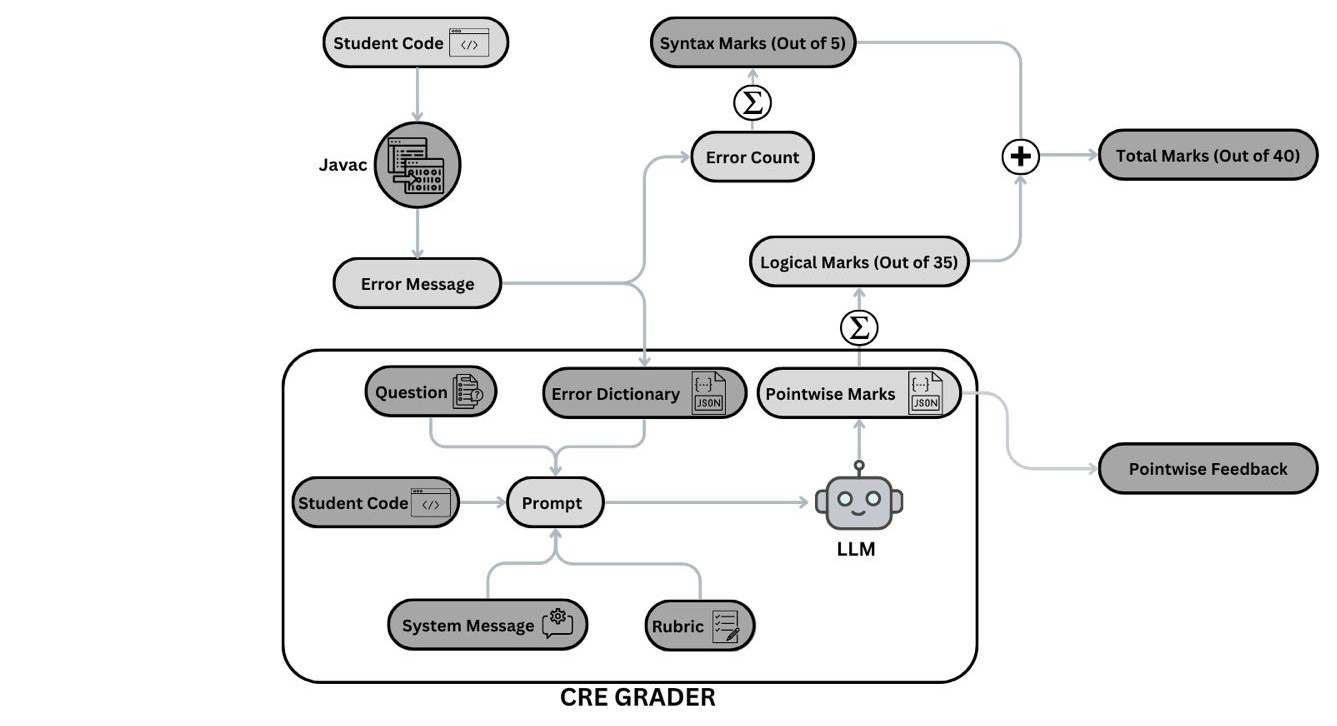}
    \caption{Complete Rubric Evaluation (CRE)}
    \label{fig:cre_flowchart}
\end{figure*}

Complete Rubric Evaluation (CRE) is a system of rubric-based evaluation developed as part of our research into LLMs and their capabilities in code evaluation. As shown in \textbf{Figure \ref{fig:cre_flowchart}}, the CRE LLM grader agent takes as input the complete problem description followed by a complete rubric (a multi-tiered point-by-point marking scheme). The rubric points act as anchors around which the student code is evaluated. Finally, the grader agent takes in the entire student code file, including all methods and classes implemented by the student.

The grader returns a single JSON dictionary without additional text. The JSON dictionary is a nested rubric, where the primary keys are method names. The values may either be the marks assigned to the method or another dictionary containing finer evaluation points for the method. Since LLMs are unreliable in arithmetic operations, the final calculation of marks is performed by a recursive function outside the LLM loop.

The LLM in CRE grader is instructed to ignore syntax errors in the student code, treating them as correct. CRE aims to infer the logical intention behind student code and mark it accordingly. This simulates a university-like evaluation environment, where logical reasoning is prioritized over minor syntax errors. This approach addresses gaps in traditional evaluation methods (e.g., test-case-based evaluation), which assign zero marks to submissions with minor errors.

While the LLM performs only logical evaluation, syntax evaluation is conducted separately using a deterministic compiler-equipped agent. The student code is executed by the agent using a compiler via a system call in a Python script. The compiler returns a syntax assessment, and the agent uses a penalty-based system to assign syntax marks accordingly. For instance, with a maximum of 5 syntax marks and a penalty of 0.5, a code snippet with 5 syntax errors is awarded 2.5 marks. The final student marks are the sum of logical and syntactical scores. Prompts used for instructing LLMs are shared in \autoref{appendix:prompts}.

\subsection{\textbf{Pointwise Rubric Evaluation (PRE)}}

Pointwise Rubric Evaluation (PRE) is a modified version of CRE. Instead of evaluating the entire rubric at once, the PRE LLM grader agent assesses the student code based on a single rubric point at a time. PRE Grader takes as input the problem statement, student code, and a single rubric point for evaluation. It returns a JSON dictionary containing the rubric point as the key and the assigned LLM marks as the value. These results are stored in a JSON file for later computations. PRE is resource-intensive due to multiple API calls per student solution. An LLM call is executed in order to evaluate each point in the rubric, consuming tokens and time.


\subsection{\textbf{Ensembling Method Evaluation (EME)}}
\begin{figure*}
    \centering
    \includegraphics[width=\linewidth]{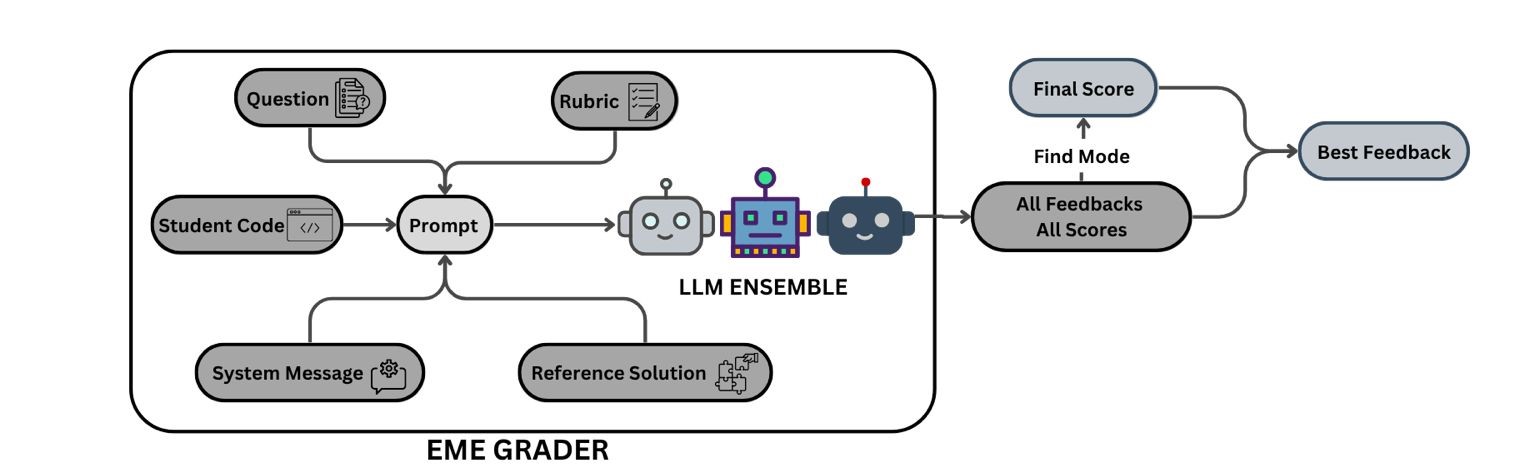}
    \caption{Ensembling Method Evaluation (EME)}
    \label{fig:eme_flowchart}
\end{figure*}

Inspired by CodEv \cite{Tseng_2024}, Ensembling Method Evaluation (EME) leverages large language models (LLMs), such as GPT-4o, Claude 3.7 Sonnet \cite{claude3.7sonnet} and GPT-4o mini, to validate ensemble-generated results using a structured evaluation framework. The method employs a sampling and voting-based approach, primarily relying on the majority voting method to determine the final ensemble output. In cases where no clear majority emerges, the rounded mean method is used as an alternative to aggregate the scores effectively. Additionally, EME incorporates a feedback selection mechanism, where the system identifies the most representative feedback by selecting the evaluation feedback whose total score is closest to the final ensemble score.
As shown in \textbf{Figure \ref{fig:eme_flowchart}}, this technique takes the question, the student solution, the reference solution and the rubric. In case of DSA dataset we also add a approach identification prompt which uses the GPT 4o model, to identify the which approach the student has used from the rubric. This prompt provides us with a confidence interval for the approach identified by the model.

%% file: files/04-evaluation.tex
\subsection{LLM selection}
In order to maintain a balance between LLM accuracy and token costs, we used OpenAI's GPT-4o-mini model to conduct our evaluations. 
We also used Anthropic AI's Claude 3.7 Sonnet model with Extended Thinking deactivated.

\subsection{Data preprocessing}

\subsubsection{Scaling}
With varying scales and degrees of rubrics, the scale at which grades are awarded is different for every technique. Here, scale refers to the maximum marks or score that can be awarded for the evaluation technique
.
In the OOP dataset, the question-specific rubric awards up to 35 points, whereas the question-agnostic rubric compresses the same outcome space into a holistic 5-point scale. Moreover, when evaluated method-by-method, a question-agnostic rubric evaluates each method on the same 0-4 scale, whereas in the original rubric, each method holds different weights in terms of maximum marks that can be awarded for that method. Therefore, in order to make quantitative comparisons between the two datasets, we scale the grades awarded by evaluation on the question-agnostic rubric to match the 35 mark question-specific rubric design.
For the purposes of understanding, we shall refer to the Ground Truth values as the Base Dataset (\textit{B}) and the LLM evaluations as the Experiment Dataset (\textit{E}). \(B\) and \(E\) contain \(N\) data points, one for each student being included in the assessment. \(B[i]\) and \(E[i]\) denote the base and experimental marks obtained by the \(i^{th}\) student, where \(0\le i<N\).
We use the terms "Base" and "Experiment" dataset with the belief that the metrics and techniques used for comparing the grading performances of an LLM grader vs a Human grader can be used to confirm the degree of affirmation between any two graders regardless of their nature, where the Experiment Grader is compared to a Base Grader in order to assess agreement.
We note that for evaluation on a method level (as seen in PRE), the experiment dataset \(E\) is two dimensional, where for each student \(i\), \(E[i]\) contains \(M\) data points, where \(M\) is the number of methods being evaluated, and \(E[i][m]\) denotes the marks obtained by student \(i\) for the method \(m\).
To compute metrics for an Experiment Dataset \textit{E} that lies on a different scale versus \textit{B}, we scale \textit{E} linearly as
\[
E_{s}[i] = E[i] \times \frac{R}{R_{E}},  \qquad 0 \leq i < N
\]
where \(R_{E}\) (e.g.\ 4 for the question-agnostic rubric) is the scale of the rubric being used for the evaluation of experiment marks, \(R\) is the scale of the question-specific rubric (also the scale of ground-truth/base values, e.g.\ 35 for the OOP dataset), \(N\) is the number of data points in the dataset, and \(E_{s}\) is the scaled experimental dataset.

For techniques evaluated method-by-method on the OOP dataset using question-agnostic rubrics, we scaled \(E\) on the method level as
\[
E_{s}[i] = \sum_{m=1}^{M} \bigl(E[i][m] \times \tfrac{R[m]}{R_{E}}\bigr), \qquad 0 \leq i < N
\]
where \(m\) is the method being evaluated, \(M\) is the number of methods present in the problem statement, and \(R[m]\) is the maximum number of marks that can be awarded for method \(m\) in the question-specific rubric. The total marks for a student are therefore the sum of marks awarded for each method, scaled up to match the base rubric.

The scaled experiment dataset \(E_s\) therefore eliminates the dimensionality issue presented by evaluating the experiment dataset \(E\) on a method level, by reducing it to a one dimensional dataset.
\subsubsection{Binning}
Before we obtain Cohen–Kappa scores on continuous datasets, we first classify the data into 5 bins. The bins are decided based on the ranks of the data points in the sorted datasets. For instance, the dataset \([3, 2, 4, 3, 1, 9, 6, 7]\) when classified into 3 bins is graded as \([0, 0, 1, 1, 0, 2, 1, 2]\), where 0, 1 and 2 are the three possible grades, assuming uniform grading. Concretely, each scaled score list is sorted and partitioned at the 20\textsuperscript{th}, 40\textsuperscript{th}, 60\textsuperscript{th} and 80\textsuperscript{th} percentiles, producing five ordinal intervals of roughly equal size.  Marks falling below a cut-off receive the lower bin label (0–4).  We adopt five bins because (i) it parallels the familiar A–E grading scale used in educational studies, (ii) it maintains adequate cell counts for stable \(\kappa\) estimates with \(N\!\approx\!100\), and (iii) sensitivity checks with 4 and 6 bins shifted \(\kappa\) by no more than 0.02.  We acknowledge that any discretisation can distort near-threshold scores, yet the empirical drift observed here is minor and does not affect qualitative conclusions.

\subsection{Evaluation Metrics}\label{sec:evaluation-metrics}

We assess each rubric--LLM pair with seven complementary statistics that together expose ordering, bias and exact‐score agreement. Although several of the following statistics are mathematically related, no single number captures \emph{all} facets of grading quality. These seven metrics can be categorized into three broad dimensions:   

\paragraph{\textbf{(1) Relative ordering of students (correlations) Rank correlations:}}\ \textbf{Pearson Correlation Coefficient} (PCC, $r$) \cite{scipy_2024} captures linear alignment and is most sensitive to large outliers.  \textbf{Spearman Rank Correlation Coefficient} ($r_s$) \cite{scipy_2024} tests whether the overall ranking is preserved regardless of spacing, while \textbf{Kendall-Tau Correlation Coefficient} ($\tau_b$) \cite{scipy_2024} provides a tie‐aware alternative that remains stable on small samples. High cross-correlation among some measures (e.g. Pearson vs. Spearman) signals convergent validity, while the
divergences highlight specific failure modes such as monotonic but non-linear trends or category-level disagreement
after binning—that would be invisible in a single score.

\paragraph{\textbf{(2) Agreement coefficients:}}  \textbf{Intraclass Correlation Coefficient (ICC)} \cite{pingouin_2024} comes in three flavours: ICC1 (one‑way random), ICC2 (two‑way random) and ICC3 (two‑way mixed). These differ in how they treat the raters (randomly drawn versus fixed).  All express absolute agreement in the original mark scale.  \textbf{Cohen-Kappa ($\kappa_B$)} \cite{sklearn_2024} treats scores as categories (bins), correcting for chance agreement; useful when instructors ultimately issue letter grades.

\paragraph{\textbf{(3) Absolute bias or strictness:}} \textbf{Leniency (Mean Normalized Error, $l_n$))} between two datasets is defined as \[l_n = \frac{\sum_{i=0}^{N-1}(\hat{E}[i]-\hat{B}[i])}{N},\]
    where \(\hat{E}[i]\) and \(\hat{B}[i]\) are the experimental and base scores normalised to 1. Leniency is a measure of how strict the evaluation system is for a rubric. Leniency is relative, and requires base data against which the metric is computed.  A perfectly lenient grader would assign \(R\) marks where every base data point is 0, achieving a leniency score of 1. A perfectly strict grader would assign 0 marks where every base data point is \(R\), achieving a leniency score of –1. If we consider a target average deviation of 10\%, the leniency to aim for would be within \(\pm 0.1\).

\paragraph{\textbf{Why Leniency matters?}}  Correlation metrics (Pearson, Spearman, Kendall) reward graders that preserve the student ranking even when every mark is shifted by a constant amount. Leniency exposes that uniform bias: a positive $l_n$ signals generosity, a negative $l_n$ strictness. Considering both together separates disagreements caused by level shifts (bias) from those caused by rank dispersion. In practice we sometimes observe $r>0.9$ while $|l_n|$ \(\pm 0.15\), indicating that a simple post‑hoc rescaling (not rubric redesign) can reconcile the two graders.

%% file: files/04a-results.tex
\subsection{Techniques Used}
\begin{itemize}[leftmargin=*]
 \item \textbf{CodeBERTScore} \cite{zhou2023codebertscoreevaluatingcodegeneration}: CodeBERTScore is an Automatic Evaluation Metric for Code evaluation, based on BERTScore. We evaluate CodeBERTScore on the OOP dataset and present correlation metrics. Leniency for CodeBERTScore is not evaluated, since the data points lie on a different scale versus the ground truth score.

 \item \textbf{CodeJudge} \cite{tong2024codejudgeevaluatingcodegeneration}: CodeJudge is an LLM-based evaluation technique . We implement the binary evaluation module in CodeJudge and evaluate binary scoring for each method. Once the score is obtained, we evaluate complete or zero marks for that method based on the marks in our rubric. CodeJudge achieves comparable results to those presented in the original paper.

 \item \textbf{Ensembling Method Evaluation (QA/QS)}: We use LLM ensembling to achieve results with a degree of consensus. Similar to CodeJudge, we evaluate each method in the original rubric, but instead of binary marking, we grade based on a 5 point rubric, awarding scores from 0-4. These marks are then scaled to 35 before evaluations. 

 \item \textbf{CRE/PRE}: Complete and Pointwise Rubric Evaluation are question-specific rubric evaluation techniques discussed earlier in the paper. CRE and PRE are both evaluated on the original 35-mark rubric. For evaluation metrics on OOP, we consider only logical marks, as human grading of the OOP dataset followed a similar logical correctness-based approach.

 \item \textbf{Five Point Marking (FPM)} \cite{phung2023generativeaiprogrammingeducation}: Five Point Marking is a question-agnostic LLM-based rubric marking technique. The student code is evaluated out of 100 split into the following categories:
        \begin{itemize}
            \item Program Format (10 Marks)
            \item Time Complexity (15 Marks)
            \item Space Complexity (15 Marks)
            \item Correctness General (30 Marks)
            \item Correctness Edge Cases (30 Marks)
        \end{itemize}
\end{itemize}
        
The scores are scaled down to 35 for comparison with ground truth values.

\subsection{Findings}
Tables 2 and 3 provide an overview of the performance of all implemented techniques on the OOP and DSA datasets.

\begin{table*}
    \centering
    \begin{tabular}{lcccccccc} 
          Method&  \(r\)&  \(r_s\)&  \(\tau_b\)&  \(l_n\)&  ICC1&  ICC2&  ICC3& \(\kappa_B\)\\ 
          \hline
          \multicolumn{9}{c}{\cellcolor{gray!15}No LLM (Similarity Based)}\\ 
          \hline
          CodeBERTScore&  0.354&  0.482&  0.343&  -&  -&  -&  -& 0.241\\ 
          \hline
          \multicolumn{9}{c}{\cellcolor{gray!15}No Rubric}\\
          \hline
          CodeJudge&  0.717&  0.745&  0.597&  -0.233&  0.479&  0.541&  0.712& 0.433\\
          \hline
          \multicolumn{9}{c}{\cellcolor{gray!15}Question-Agnostic (QA) Rubric}\\
          \hline
          EME (QA)&  0.904&   \textbf{0.909}&  \textbf{0.775}&   -0.071&  \textbf{0.881}&  \textbf{0.882}&  0.904& 0.512\\ 
          FPM&  0.844&  0.834&  0.670&  -0.121&  0.778&  0.784&  0.834& 0.346\\
          \hline
 \multicolumn{9}{c}{\cellcolor{gray!15}Question-Specific (QS) Rubric}\\
 \hline
 EME (QS)& 0.900& 0.902& 0.769& -0.067& {0.878}& {0.880}& 0.900&0.545\\ 
 CRE (OpenAI)& \textbf{0.912}& 0.906& 0.773& 0.082& 0.880& 0.882& \textbf{0.910}&\textbf{0.598}\\
 CRE (Claude)& 0.840& 0.841& 0.696& \textbf{-0.008}& 0.841& 0.841& 0.840&0.572\\ 
 PRE& 0.742& 0.795& 0.637& -0.329& 0.201& 0.378& 0.679&0.302\\ 
    \end{tabular}
    \caption{Results obtained by various techniques on the OOP dataset}
    \label{tab:oop_results}
\end{table*}

\begin{table*}
    \centering
    \begin{tabular}{lcccccccc} 
         Method&  \(r\)&  \(r_s\)&  \(\tau_b\)&  \(l_n\)&  ICC1&  ICC2&  ICC3& \(\kappa_B\)\\
         \hline
 \multicolumn{9}{c}{\cellcolor{gray!15}No LLM}\\
 \hline
 CodeBERTScore& 0.126& 0.058& 0.039& -& -& -& -&0.010\\
 \hline
 \multicolumn{9}{c}{\cellcolor{gray!15}No Rubric}\\
 \hline
 CodeJudge& 0.423& 0.427& 0.389& -0.315& 0.176& 0.272& 0.353&0.406\\
\hline
         \multicolumn{9}{c}{\cellcolor{gray!15}Question-Agnostic (QA) Rubric}\\ 
         \hline
         EME (QA)&  0.562&  0.510&  0.445&  -0.098&  0.509&  0.525& 0.560& 0.156\\
 FPM& 0.470& 0.381& 0.320& -0.054& 0.380& 0.388& 0.398&0.072\\
 \hline
 \multicolumn{9}{c}{\cellcolor{gray!15}Question-Specific (QS) Rubric}\\ 
 \hline
         EME (QS)&  \textbf{0.825}&  \textbf{0.763}&  \textbf{0.675}&  \textbf{0.0049}&  \textbf{0.821}&  \textbf{0.820}&  \textbf{0.819}& \textbf{0.646}\\ 
    \end{tabular}
    \caption{Results obtained by various techniques on the DSA dataset}
    \label{tab:dsa_results}
\end{table*}


\subsubsection{Overall Observations (\autoref{tab:oop_results} and \autoref{tab:dsa_results})}
\begin{itemize}
    \item \textbf{Presence of LLM Grader}: We observe that LLM-based techniques outperform CodeBERTScore significantly and
impressively. CodeBERTScore obtains weak scores across all evaluation metrics. LLMs are vastly more versatile when it
comes to understanding context, as well as the varying nature of student approaches to one particular problem.
\item \textbf{Presence of Rubric:} We observe that both question-agnostic and question-specific rubric techniques outperform
no-rubric technique (CodeJudge). While CodeJudge achieves respectable correlation and \(\kappa_B\) scores, providing a rubric gives the
LLM grader an anchor around which to evaluate or grade the code, thereby increasing performance significantly when
a rubric is provided
\end{itemize}
\subsubsection{Specific Observations from Results on DSA Dataset (\autoref{tab:dsa_results})}
\begin{itemize}
  \item \textbf{Question‐specific beats question‐agnostic rubric-based approaches:}
        On the algorithmically diverse and hard \textsc{DSA} dataset, moving from a question‐agnostic rubric \emph{(EME‐QA)} to a question‐specific rubric \emph{(EME‐QS)} lifts $\mathrm{ICC3}$ from $0.560 \rightarrow 0.819$ and boosts Pearson $r$ by $+0.26$ points $(0.562 \rightarrow 0.825)$.
        \item   \textbf{Type of Question-Agnostic rubric:} Using EME with a QA rubric provides significantly better correlation scores versus the FPM technique. The FPM technique attempts to grade the solution over 5 predefined marking points. An LLM when provided with only a problem statement and a solution code snippet is unable to grade accurately based on these 5 marking points, thereby yielding lower scores. In contrast, it performs better when asked to gauge the degree of correctness of the entire code snippet directly.

\end{itemize}
\subsubsection{Specific Observations from OOP Results (\autoref{tab:oop_results})}
\begin{itemize}
    \item \textbf{Both question-specific and question-agnostic rubrics achieve comparable results for EME:}
        We see comparable performance for question-specific (EMA(QS)) and question-agnostic (EMA(QA)) rubric-based approaches across all metrics. As the OOP dataset contains homogeneous implementation-oriented questions, the results indicate that such questions are graded reliably even with question-agnostic rubrics.
    
    \item   \textbf{Strictness depends on prompt granularity:}
        As shown in \autoref{tab:oop_results}, \emph{PRE} (feeding one rubric point at a time) slashes average scores by $11.5/35$ marks (leniency $=-0.329$), whereas \emph{CRE} (feeding the whole rubric) hovers near human leniency $(0.081)$.
        \item   \textbf{Why PRE is harsher than CRE?}
        Single-criterion prompts force the model to assign zero unless the exact logic is present, whereas whole-rubric prompts let it award partial credit, mirroring human evaluator behaviour. PRE seems to be more suitable for use cases requiring stringent evaluation and strict rubric adherence.
\end{itemize}
\subsubsection{EME Performance}
We conducted an additional set of experiments to understand the impact of model parameter size and ensemble size for EME.
\begin{itemize}
    \item \textbf{Model Parameter Size}: The performance of EME improves as the parameter size of the underlying language model increases. Larger models tend to demonstrate enhanced reasoning and evaluation capabilities, resulting in more accurate scoring and feedback generation. Conversely, models with relatively lower parameter counts exhibit greater inconsistencies, leading to lower correlation.

    \item \textbf{Ensemble Size}: The effectiveness of the method varies with ensemble size. Initially, a significant increase in correlation coefficients is observed as ensemble size increases. This trend continues until the ensemble reaches an optimal range—typically around three to four models when employing high-parameter LLMs, such as GPT-4 and Claude 3. Beyond this threshold, performance gains tend to plateau, suggesting diminishing returns with further increase in ensemble size.
\end{itemize}

%% file: files/06b-discussion.tex
\subsection{Implications for Instructors and Teaching Assistants}

For instructors and teaching assistants, the three graders (CRE, PRE and EME) can be combined in a staged workflow. A lightweight \textsc{CRE} pass may quickly triage submissions, flagging clearly strong or weak attempts, whereas stricter \textsc{PRE} scoring (or a brief human check) can be reserved for borderline cases. In settings where false positives carry a high cost, such as summative examinations, \textsc{PRE} (strict) may offer additional reassurance, trading a slight reduction in overall agreement for a zero‑tolerance stance on partial logic. Routine coursework and large MOOC cohorts, by contrast, often prioritise speed and cost. For such scenarios, \textsc{CRE} 
seems to be a reasonable cost-effective choice.
Classes that attract highly diverse algorithmic solutions (for example, DSA) could find value in \textsc{EME}~(QS), which attained the highest ICC3 in our \textsc{DSA} benchmark, albeit at the price of a three‑model ensemble.
We also note that for larger, easier and more direct problems, as seen in the OOP dataset, a simple question-agnostic rubric with method-wise evaluation produces near-human-like grading and is on par with evaluation using question-specific rubrics. It does, however, rely on multiple LLM calls (once for each method) rather than a single LLM call with a large rubric. Developing a question-specific rubric may therefore save token costs during evaluation.

The JSON traces emitted by all three techniques may also feed into analytics dashboards. Aggregating the most frequently missed rubric criteria can highlight common misconceptions, enabling teaching teams to design targeted recitations, micro‑lectures, or discussion‑board posts that address exactly those weak spots.

A minimal integration path could involve: (i) translating an existing rubric into the key–value format accepted by the graders, (ii) calibrating Leniency or ensemble size on roughly ten pilot submissions until the mean‑normalised error falls within $\pm0.1$, (iii) inserting the grader call into the continuous‑integration script that already compiles and tests student code, and (iv) offering reviewers an interface that surfaces rubric items with confidence below 0.8 for optional human override.

\subsection{Guidance for Students}

Rubric‑aligned feedback provides criterion‑level transparency: each comment is anchored to a specific step (e.g., Step 6 – create a \texttt{Player} object), helping learners understand \emph{why} marks were lost instead of receiving a generic wrong output notice. Encouraging students to summarise the feedback in a simple worksheet (criterion missed, probable cause, planned fix) may promote systematic debugging. Learners might further consolidate gains by submitting a short reflection after resubmission detailing which rubric elements they have mastered, which remain challenging, and what strategies (additional unit tests, peer review, etc.) they will employ next time. Such reflective practice aligns with accepted principles of self‑regulated learning.

%

%% file: files/06a-limitations.tex
Although this study provides valuable insights into rubric-based code evaluation, certain limitations must be acknowledged which also advocate avenues for future work. 
Firstly, we didn't explore variability among various LLMs and mainly worked with GPT-4o. Different training methodologies may influence effectiveness and accuracy of code evaluation and a future research could compare multiple models. We also didn't explore the differences between open-sources and closed-source LLMs.
Our analysis is limited to programming questions in Java and do not test other widely-used languages like Python, C++. 
We focussed on two intermediary courses which had single file code assignments. A future research could delve into advanced courses which require multiple files handling. The rubrics can be of various granularities ranging from low to medium to fine. A detailed study on their effectiveness and comparison is another future avenue for research. 

%% file: files/06-conclusion.tex
In this paper, we explored how effectively can an LLM-based grader evaluate student code using a question-specific rubric compared to a question-agnostic rubric. 
To facilitate our evaluation, we introduced two novel datasets focused on DSA and OOP, encompassing solutions of varying correctness levels. Using these datasets, we assessed our proposed techniques against existing approaches using exisitng and a new evaluation metric, Leniency.
Our findings highlight the limitations of existing evaluation techniques and demonstrate the potential benefits of question-specific rubrics in enhancing logical assessment of code. This work not only provides new insights into LLM-based code evaluation but also lays the groundwork for future research into refining evaluation methods and expanding dataset availability to improve automated code assessment.

%% file: files/061-acknowledgements.tex
This research was carried out in part with support from the New Faculty Seed Grant, Birla Institute of Technology and Science (BITS), Pilani (Grant Ref. N4/24/1004). 

The authors also wish to acknowledge the use of ChatGPT/Claude in the writing of this paper. This tool was used to generate ideas regarding the presentation of tables and figures in the paper, and to improve the written grammar. The paper remains an accurate representation of the authors’ underlying work and novel intellectual contributions.

%% file: files/07-appendix.tex
\section{Prompts for Techniques}\label{appendix:prompts}
\subsection{CRE}
\begin{lstlisting}
You are an expert code evaluator, evaluating code submissions for a Java based Object Oriented Programming test at a university level.
You will be provided with the question and a rubric that describes the criteria for evaluation, with a marking scheme. 
The question is a code sample that the examiner provides, containing a template wherein the student is required to write the code as well as comments and instructions from the examiner's end.
Following this you will be provided with the code submission, along with the response from the Java compiler that runs this code.
Note that the code may be formatted liberally, the specific positioning of the code within the methods are not important.
Code may be present either before or after the comments prepared by the instructor.
You are to evaluate the code based only on logical correctness. You are to ignore any syntax errors that the compiler may have thrown. 
Any syntax errors that you encounter can be treated as correct syntax, and you are to infer the student's logical flow and intention from the code.
You are to return your response as a JSON dictionary containing a detailed, nested evaluation of the student's marks for each line in the rubric.
The JSON dictionary should also contain feedback for each point in the rubric.
For each line in the rubric, you are to provide the line as the key and a nested dictionary containing marks awarded and feedback.
The following is a sample return Format:
{
"1000": {
    "Method1": {
        "Point1": {
            "Marks": 3,
            "Feedback": {LLM obtained feedback}
        },
        "Point2":{...},
        ...
    },
    "Method2: {...},
    ...
}
}
DO NOT RETURN ANY ADDITIONAL TEXT ASIDE FROM THE JSON DICTIONARY.
Question: {}
Rubric: {}
Code Submission: {}
Compiler Response: {}
\end{lstlisting}

\subsection{PRE}
\begin{lstlisting}
You are an expert code evaluator, evaluating code submissions for a Java based Object Oriented Programming test at a university level.
You will be provided with the question, the code snippet, and the point of evaluation for the code. You will also be given the compiler response for the code.
You will also be given the rubric point that the student is graded on. You are to evaluate based on that particular point only.
The question is a code sample that the examiner provides, containing a template wherein the student is required to write the code as well as comments and instructions from the examiner's end.
Following this you will be provided with the code submission, along with the response from the Java compiler that runs this code.
Note that the code may be formatted liberally, the specific positioning of the code within the methods are not important.
Code may be present either before or after the comments prepared by the instructor.
You are to evaluate the code based only on logical correctness. You are to ignore any syntax errors that the compiler may throw. 
Any syntax errors that you encounter can be treated as correct syntax, and you are to infer the student's logical flow and intention from the code.
You are to return only a dictionary containing the your decision and your feedback, with the keys "DECISION" and "FEEDBACK"
For your decision, return YES if the student has correctly implemented the logic for the given rubric point, and NO if they have not.
Since there is no partial marking and we're only considering logical correctness, be liberal with the quality of the code and the marking.
The following is a sample return Format:
{
    "DECISION": "YES",
    "FEEDBACK": {LLM obtained Feedback},
}
DO NOT RETURN ANY ADDITIONAL TEXT ASIDE FROM THE DICTONARY.
Question: {}
Student Solution: {}
Point to be evaluated: {}
Compiler Response: {}
\end{lstlisting}
\subsection{EME}
\begin{lstlisting}
## Approach Identification Prompt

You are analyzing a student's code submission for a DSA problem.
Based on the rubric, identify which approach the student is using.

Rubric:
```
{rubric_content}
```

Student Code:
```
{code}
```

IMPORTANT INSTRUCTIONS:
1. The rubric contains multiple solution approaches (e.g., "Solution 1", "Solution 2", "Solution 3").
2. Each approach has specific criteria and point allocations.
3. Carefully analyze the student's code to determine which approach they are using.
4. Look for key patterns, variable names, and algorithm structures that match one of the approaches in the rubric.
5. The approach name should be EXACTLY as it appears in the rubric (e.g., "Brute Force", "Dynamic Programming", "Kadane's Algorithm").

Respond ONLY with a JSON object in this exact format:
{
    "identified_approach": "Exact approach name from rubric (e.g., 'Solution 1 (Brute Force)', 'Solution 3 (Kadane's Algorithm)')",
    "confidence": 0.95, // A number between 0 and 1 indicating confidence in the identification
    "reasoning": "Brief explanation of why you identified this approach, citing specific code patterns that match the rubric criteria"
}

## System Message for Approach Identification
You are a code analyzer that ONLY responds with valid JSON. No other text or explanation. You must identify the exact approach from the rubric.

## Code Evaluation Prompt

You are evaluating a student's code submission for a DSA problem.
Provide your evaluation in VALID JSON format only.

Problem:
```
{question}
```

Rubric:
```
{rubric}
```

Reference Solution:
```
{solution}
```

Student Code:
```
{code}
```

The student appears to be using the "{identified_approach}" approach.

IMPORTANT INSTRUCTIONS:
1. Evaluate the submission according to the EXACT criteria in the rubric for this approach.
2. For each criterion in the rubric for this approach, assign appropriate points.
3. The criterion descriptions should match EXACTLY what's in the rubric.
4. The max_score for each criterion should match the points specified in the rubric.
5. Your feedback should directly address how well the student's code meets each specific criterion.
6. Do not create new criteria that aren't in the rubric.

Respond ONLY with a JSON object in this exact format:
{
    "criteria_scores": [
        {"criterion": "exact criterion from rubric", "score": awarded_points, "max_score": points_specified_in_rubric, "feedback": "specific feedback for this criterion"}
    ],
    "total_score": total_awarded_points,
    "max_possible_score": total_maximum_points,
    "overall_feedback": "overall feedback here",
    "approach_correctness": 0.95, // How confident you are that the approach identification is correct (0-1)
    "code_correctness": 0.9, // How likely the code is to work correctly (0-1)
    "efficiency_rating": 0.8, // How efficient the solution is relative to optimal (0-1)
    "readability_rating": 0.7 // How readable and well-structured the code is (0-1)
}

## System Message for Code Evaluation
You are a code evaluator that ONLY responds with valid JSON. No other text or explanation. You must follow the rubric exactly when evaluating code. 
\end{lstlisting}